\documentclass[aps,showpacs,floatfix,twocolumn]{revtex4}
\usepackage{graphicx}
\usepackage{amsmath}
\usepackage{amssymb}
\usepackage{amsfonts}
\usepackage{bm}
\usepackage{color}

\def\be{\begin{equation}}
\def\ee{\end{equation}}
\def\bea{\begin{eqnarray}}
\def\eea{\end{eqnarray}}

\begin{document}

\title{Gravitational collapse of Bose-Einstein condensate dark matter halos}

\author{Tiberiu Harko}
\email{t.harko@ucl.ac.uk}
\affiliation{Department of Mathematics, University College London, Gower Street, London
WC1E 6BT, United Kingdom}

\date{\today}

\begin{abstract}
We study the mechanisms of the gravitational collapse of the Bose-Einstein condensate dark matter halos, described by the zero temperature time-dependent nonlinear Schr\"odinger equation (the Gross-Pitaevskii equation),  with repulsive inter-particle interactions.  By using a variational approach, and by choosing an appropriate trial wave function, we reformulate the Gross-Pitaevskii equation with spherical symmetry as Newton's equation of motion for a particle in an effective potential, which is determined by the zero point kinetic energy, the gravitational energy, and the particles interaction energy, respectively. The velocity of the condensate is proportional to the radial distance, with a time dependent proportionality function. The equation of motion of the collapsing dark matter condensate is studied by using both analytical and numerical methods. The collapse of the condensate ends with the formation of a stable configuration, corresponding to the minimum of the effective potential. The radius and the mass of the resulting dark matter  object are obtained, as well as the collapse time of the condensate. The numerical values of these global astrophysical quantities, characterizing condensed dark matter systems, strongly depend on the two parameters describing the condensate,  the mass of the dark matter particle, and of the scattering length, respectively.  The stability of the condensate under small perturbations is also studied, and the oscillations frequency of the halo is obtained. Hence these results show that  the gravitational collapse of the condensed dark matter halos can lead to the formation of stable astrophysical systems with both galactic and stellar sizes.
\end{abstract}


\maketitle

\section{Introduction}

The recently published Planck satellite data \cite{Pl} have generally confirmed the predictions of the standard $\Lambda $CDM ($\Lambda $Cold Dark Matter) cosmological model, as well as the matter composition of the Universe. The $\Lambda $CDM
model successfully describes the accelerated expansion of the Universe, the
observed temperature fluctuations in the cosmic microwave background
radiation, the large scale matter distribution, and the main aspects of the
formation and the evolution of virialized cosmological objects. On the other hand the latest Cosmic Microwave Background (CMB) data, as well as the observations of the distant Type IA  supernovae,  baryon acoustic oscillations (BAO), weak gravitational lensing, and the abundance of galaxy clusters,  provide compelling evidence that about 95\% of the content of the Universe resides in two unknown forms of matter/energy, called
dark matter and dark energy, respectively: the first residing in bound objects as
non-luminous matter at the galactic and extragalactic scale \cite{dm}, while the latter is in the form of a zero-point energy that
pervades the whole Universe \cite{Rev0, PeRa03}. The dark matter is assumed to be
composed of cold neutral weakly interacting massive particles, beyond those
existing in the Standard Model of Particle Physics, and not yet detected in
accelerators or in dedicated direct and indirect searches. There are many
possible candidates for dark matter, the most popular ones being the axions
and the weakly interacting massive particles (WIMP) (for a review of the
particle physics aspects of dark matter see \cite{OvWe04}). The
interaction cross sections of dark matter particles with normal baryonic matter, while extremely
small, are expected to be non-zero, and we may expect to detect them
directly \cite{AMS}. Scalar fields or other long range
coherent fields coupled to gravity have also intensively been used to model
galactic dark matter \cite{scal}. Alternative
theoretical models to explain the galactic rotation curves have also
been elaborated recently \cite{alt}.

Despite its important achievements, at galactic scales of the order of $\sim 10$ kpc, the $%
\Lambda $CDM model faces major challenges in explaining the observed
distribution of the dark matter around the luminous one. In fact, $N$%
-body simulations, performed in the $\Lambda $CDM scenario, have shown that bound halos
surrounding galaxies must have very characteristic density profiles that
feature a well pronounced central cusp, $\rho
_{NFW}(r)=\rho _{s}/(r/r_{s})(1+r/r_{s})^{2}$ \cite{nfw}, where $r_{s}$ is a scale
radius and $\rho _{s}$ is a characteristic density. On the observational
side, high-resolution rotation curves show, instead, that the actual
distribution of dark matter is much shallower than the simulated one, and it
presents a nearly constant density core: $\rho _{B}(r)=\rho
_{0}r_{0}^{3}/(r+r_{0})(r^{2}+r_{0}^{2})$ \cite{bur}, where $r_{0}$ is the core radius
and $\rho _{0}$ is the central density. Therefore, to solve this contradiction between observation and theory, new models and a new understanding of dark matter and its properties are required.

The observation of the Bose-Einstein condensation in 1995 in dilute alkali gases, such as vapors of rubidium and sodium, confined
in a magnetic trap, and cooled to very low temperatures \cite{exp} did represent a major breakthrough in condensed matter and statistical physics. At very low temperatures, all particles in a dilute Bose gas condense to the
same quantum ground state, forming a Bose-Einstein Condensate (BEC), i.e., a sharp peak over a broader
distribution in both coordinates and momentum space. Particles become
correlated with each other when their wavelengths overlap, that is, the
thermal wavelength $\lambda _{T}$ is greater than the mean inter-particles
distance $l$. This happens at a temperature $T<2\pi \hbar ^{2}n^{2/3}/mk_{B}$%
, where $m$ is the mass of the particle in the condensate, $n$ is the number
density, and $k_{B}$ is Boltzmann's constant \cite{Da99, rev,Pit,Pet,Zar}. A coherent state
develops when the particle density is  high enough, or the temperature is
sufficiently low.  From experimental point of view, the occurrence of the condensation is indicated by a sharp peak in the
velocity distribution, observed below a critical temperature. This show that all the atoms have condensed in the same
ground state, with a narrow peak in the momentum  and
coordinate space \cite{exp}.
Quantum degenerate gases have been created by a
combination of laser and evaporative cooling techniques, opening several new
lines of research, at the border of atomic, statistical and condensed matter
physics \cite{Da99,rev,Pit,Pet,Zar}.

Since the Bose-Einstein condensation is a phenomenon observed and well studied in the laboratory, the possibility that it may occur on astrophysical or cosmic scales cannot be rejected {\it a priori}. Thus, dark matter,  which is required to explain the
dynamics of the neutral hydrogen clouds at large distances from the galactic
center, and which is a cold, bosonic system, could also be in the form of a Bose-Einstein condensate \cite{Sin}. In this early studies either a phenomenological approach was used, or  the
 non-relativistic Gross-Pitaevskii equation describing the condensate  was investigated numerically. A systematic study of the condensed galactic dark matter halos and of their properties was performed in \cite{BoHa07}. By
introducing the Madelung representation of the wave function, the dynamics
of the  dark matter halo can be formulated in terms of the continuity equation and of
the hydrodynamic Euler equations. Hence condensed dark matter can be described as a
non-relativistic, Newtonian  gas,
whose density and pressure are related by a barotropic equation of state. In
the case of a condensate with quartic non-linearity, the equation of state
is polytropic with index $n=1$ \cite{BoHa07}.

To test the validity of the condensed dark matter model  the
Newtonian tangential velocity equation  was fitted with a sample
of rotation curves of low surface brightness and dwarf galaxies,
respectively. A very good agreement was found between the theoretical
rotation curves and the observational data.  Therefore dark matter halos can be described
as  an assembly of light individual bosons that acquire a repulsive interaction by
occupying the same ground energy state. The repulsive interaction prevents gravity from forming
the central density cusps. The condensate particle is light enough to
naturally form condensates of very small masses that later may coalesce,
forming the structures of the Universe in a similar way than the
hierarchical clustering of the bottom-up CDM picture. Then, at large scales,
BEC perfectly mimic an ensemble of cold particles, while at small scales
quantum mechanics drives the mass distribution.

The properties of the Bose-Einstein condensed dark matter halos, as well as their cosmological implications,  have been intensively investigate recently.  The recently observed size evolution of very massive compact galaxies in the early universe can be explained, if dark matter is in a Bose Einstein Condensate state \cite{Lee0}. The size of the dark matter halos and galaxies depends on the correlation length of dark matter and, hence, on  the expansion of the universe. The BEC predicts that the size of the galaxies increases as the Hubble radius of the universe even without merging, which agrees well with the recent observational data. In \cite{Lee1}  it was shown that the finite length scale of the condensate dark matter can  explain the recently observed common central mass of the Milky Way satellites ($\sim 10^7M_{\odot}$) independent of their luminosity, if the mass of the dark matter particle is about $10^{-22}$ eV.

The validity of the BEC model on the galactic scale by using observed rotation curves was tested in \cite{Har2} by comparing  the tangential velocity equation of the model with a sample of eight rotation curves of dwarf galaxies. A good agreement was found between the theoretically predicted rotation curves (without any baryonic component) and the observational data. The mean value of the logarithmic inner slope of the mass density profile of dwarf galaxies was also obtained, and it was shown that the observed value of this parameter is in agreement with the theoretical results.
The study of the galactic rotation curves in the BEC model was considered in \cite{Mat1} and \cite{Ger}.  The BEC model predicts that all galaxies must be very similar and exist for bigger redshifts than in the $\Lambda $CDM model.  In \cite{Mat1} the fits of high-resolution rotation curves of a sample of thirteen low surface brightness galaxies were compared with fits obtained using a Navarro-Frenk-White and Pseudo-Isothermal (PI) profiles. A better agreement with the BEC model and PI profiles was found. The mean value of the logarithmic inner density slopes is -0.27 +/- 0.18. A natural way to define the core radius, with the advantage of being model-independent, was also introduced. Using this new definition in the BEC density profile it was found that the recent observation of the constant dark matter central surface density can be reproduced. The BEC model gives a similar fit to the Navarro-Frenk-White dark matter model for all High Surface Brightness (HSB) and Low Surface Brightness (LSB) galaxies in a sample of 9 galaxies \cite{Ger}. For dark matter dominated dwarf galaxies the addition of the BEC component improved more upon the purely baryonic fit than the NFW component. Thus despite the sharp cut-off of the halo density, the BEC dark matter candidate is consistent with the rotation curve data of all types of galaxies \cite{Ger}.  The dynamics of rotating Bose Condensate galactic dark matter halos, made of an ultralight spinless bosons, and the impact of the halo rotation on the galactic rotation curves was analyzed in \cite{Guz1}.

Finite temperature effects on  dark matter halos were analyzed in \cite{Har1}, where the condensed dark matter and thermal cloud density and mass profiles at finite temperatures were explicitly obtained. The obtained results show that when the temperature of the condensate and of the thermal cloud are much smaller than the critical Bose-Einstein transition temperature, the zero temperature density and mass profiles give an excellent description of the dark matter halos. The angular momentum and vortices in BEC galactic dark matter halos were studied in \cite{Kai}-\cite{Shap2}, respectively.

In \cite{Pir} it was  proposed that the dark matter content of galaxies consist of a cold bosonic fluid, composed of Weakly Interacting Slim Particles (WISPs), represented by spin-0 axion-like particles and spin-1 hidden bosons, thermalized in the Bose-Einstein condensation state and bounded by their self-gravitational potential.  By comparing this model with data obtained from 42 spiral galaxies and 19 Low Surface Brightness  galaxies,  the dark matter particle mass was constrained to the range $10^{-6}-10^{-4}$ eV, and the lower bound for the scattering length was found to be of the order of $10^{-14}$ fm.

The possibility that due to their superfluid properties some compact astrophysical objects may contain a significant part of their matter in the form of a Bose-Einstein condensate was investigated in \cite{Harko3}. To study the condensate the Gross-Pitaevskii equation was used, with arbitrary non-linearity.  In this way a large class of stable astrophysical objects was obtained, whose basic astrophysical parameters (mass and radius) sensitively depend on the mass of the condensed particle, and on the scattering length.

The Bose-Einstein condensation process in a cosmological context, by assuming that this process can be described (at least approximately) as a first order phase transition was studied in \cite{Har4}. It was shown that the presence of the condensate dark matter and of the Bose-Einstein phase transition could have modified drastically the cosmological evolution of the early universe, as well as the large scale structure formation process. The effects of the finite dark matter temperature on the properties of the Bose-Einstein Condensed dark matter halos were analyzed, in a cosmological context, in \cite{Har5}. The basic equations describing the finite temperature condensate, representing a generalized Gross-Pitaevskii equation that takes into account the presence of the thermal cloud were formulated. The static condensate and thermal cloud in thermodynamic equilibrium was analyzed in detail, by using the Hartree-Fock-Bogoliubov and Thomas-Fermi approximations.  It was also shown that finite temperature effects may play an important role in the early stages of the cosmological evolution of the dark matter condensates. The cosmological perturbations in the cosmological models with condensed dark matter were studied in \cite{Har6,Chav,Frei}.  The large scale perturbative dynamics of the BEC dark matter in a model where this component coexists with baryonic matter and cosmological constant was investigated in \cite{Wam}. The perturbative dynamics was studied using neo- Newtonian cosmology (where the pressure is dynamically relevant for the homogeneous and isotropic background) which is assumed to be correct for small values of the sound speed.  BEC dark matter effects can be seen in the matter power spectrum if the mass of the condensate particle lies in the range $15\; {\rm meV} < m < 700\; {\rm meV}$, leading to a small, but perceptible, excess of power at large scales.

Simulation codes that are designed to study the behavior of the dark matter galactic halos in the form of a Bose-Einstein Condensate were developed in \cite{Mad} and \cite{Guz2}. In \cite{Pires1} it was shown  that once  appropriate choices for the dark matter particle mass and scattering length are made, the galactic dark matter halos composed by axion-like  Bose-Einstein Condensed particles,  trapped by a self-gravitating potential, may be stable in the Thomas-Fermi approximation.  The validity of the Thomas-Fermi approximation for the halo system was also discussed, and it was shown that the kinetic energy contribution is indeed negligible. The Thomas-Fermi approximation for the study of the condensed dark matter halos was also discussed in \cite{Toth}. The Thomas-Fermi approximation is based on the assumption that in the presence of a large number of particles, the kinetic term in the Gross-Pitaevskii energy functional can be neglected. However,  this assumption is violated near the condensate surface. It was also shown that the total energy of the self-gravitating condensate in the Thomas-Fermi approximation is positive.

A major recent experimental advance in the study of the Bose-Einstein condensation processes was the observation of the collapse and subsequent explosion of the condensates \cite{Don}.  A dynamical study of an attractive
$^{85}$Rb BEC in an axially symmetric trap was done, where
the interatomic interaction was manipulated by changing the external
magnetic field, thus exploiting a nearby Feshbach resonance. In the vicinity of a Feshbach resonance the atomic scattering length a can be varied over a huge range, by adjusting
an external magnetic field. Consequently, the
sign of the scattering length is changed, thus transforming a repulsive
condensate of $^{85}$Rb atoms into an attractive one, which naturally
evolves into a collapsing and exploding condensate.

From a simple physical point of view the collapse of the Bose-Einstein Condensates can be described as follows. When the number of particles becomes sufficiently large, so that $N>N_c$, where $N_c$ is a critical number, the
attractive inter-particle energy overcomes the quantum pressure,
and the condensate implodes. In the course of the implosion
stage, the density of particles increases in the small
vicinity of the trap center. When it approaches a certain critical
value, a fraction of the particles gets expelled. In a time
period of an order of few milliseconds, the condensate again
stabilizes. There are two observable components at the final
stage of the collapse: remnant and burst particles. The remnant
particles are those which remain in the condensate. The burst
particles have an energy much larger than that of the condensed
particles. There is also a fraction of particles, which is not observable.
This fraction is referred to as the missing particles \cite{Ryb04}.

 The study of the Bose-Einstein  collapse within a model of a gas of free bosons described by a semi-classical Fokker-Planck equation was performed in \cite{Chavcol} and \cite{Chav2}. A striking similarity  between the Bose-Einstein condensation in the canonical ensemble, and the gravitational collapse of a gas of classical self-gravitating Brownian particles was found. It was also shown that at the Bose-Einstein condensation temperature $T_c$, the chemical potential $\mu (t)$ vanishes exponentially with a universal rate. After $t_{coll}$, the finite time interval in which $\sqrt{\mu (t)}$ vanishes, the mass of the condensate grows linearly with time, and saturates exponentially to its equilibrium value for large times \cite{Chavcol}.

It is the purpose of the present paper to study the dynamics of gravitationally self-bound
Bose-Einstein dark matter condensates of collisionless particles, without exterior trapping potentials. In particular, we focus on the description, mechanism and properties of the condensate collapse. In order to study the gravitational collapse, and to solve the Gross-Pitaevskii equation describing the dynamics of the condensate,  we employ a variational method. By appropriately choosing a trial wave function, the dynamical evolution of the condensate can be described by an effective time - dependent action, with the equation of motion of the condensate being given by the equation of motion of a single particle in an effective potential. The effective potential contains the contributions of the zero point kinetic energy, of the gravitational energy, and of the interaction energy, respectively. The effective equation of motion of the collapsing dark matter condensate is studied by using both analytical and numerical methods. The collapse of the condensate ends with the formation of a stable astrophysical configuration, corresponding to the minimum of the effective potential. The radius and the mass of the resulting dark matter  object are obtained, as well as the collapse time of the condensate by numerically solving the effective equation of motion. Approximate expressions for the radius of the stable configuration and of the collapse time are also obtained. The numerical values of these global astrophysical quantities, characterizing condensed dark matter systems, strongly depend on the two parameters describing the condensate,  the mass of the dark matter particle, and of the scattering length, respectively.  The stability of the condensate under small perturbations is also studied, and the oscillations frequency of the halo is obtained. Hence the results obtained in the present paper show that  the gravitational collapse of the condensed dark matter halos can lead to the formation of stable astrophysical systems on both galactic and stellar scales.

The present paper is organized as follows. The basic physical properties of the static Bose-Einstein condensed dark matter halos are reviewed briefly in Section~\ref{sect2}. The variational formulation of the Gross-Pitaevskii equation, the choice of the trial wave function, and the formulation of the effective dynamics of the condensate as a motion of a single particle in an effective potential are presented in Section~\ref{sect3}. The physical parameters of the time dependent dark matter halos (density, gravitational potential, and the physical parameters of the effective potential) are determined, within the framework of the variational approach in Section~\ref{sect4}.  The gravitational collapse of the dark matter halos is analyzed in Section~\ref{sect5}. The stability properties of the dark matter halos formed after the gravitational collapse are investigated in Section Section~\ref{sectnn}. We discuss and conclude our results in Section~\ref{sect6}.

\section{Physical parameters of the static condensed dark matter halos}\label{sect2}

In the following we assume that the dark matter halos are composed of a strongly - coupled, dilute Bose-Einstein
Condensate at absolute zero temperature. Therefore almost all the dark matter particles are
in the condensate.  In a dilute and cold gas,
only binary collisions at low energy are relevant, and, hence, independently of the details of the two-body potential, the collisions are characterized by a single physical parameter, the $s$-wave scattering length $a$. Consequently, one can replace the interaction potential with an effective
interaction $V_I\left( \vec{r}^{\;\prime }-\vec{r}\right) =u_0 \delta \left(
\vec{r}^{\;\prime }-\vec{r}\right) $, where the coupling constant $u_0 $
is related to the scattering length $a$
through $u_0 =4\pi \hbar ^{2}a/m_{\chi }$, where $m_{\chi}$ is the mass of the dark matter particle \cite{Da99}.
The ground state properties of the dark matter are
described by the mean-field Gross-Pitaevskii (GP) equation.

The density distribution $\rho _{BE}$ of the static gravitationally bounded
single component dark matter Bose-Einstein Condensate is given by \cite{BoHa07},
\be
\rho_{BE}\left( r\right) =\rho _{BE}^{(c)}\frac{\sin kr}{kr},
\ee
where $k=\sqrt{%
Gm_{\chi }^{3}/\hbar ^{2}a}$ and $\rho _{BE}^{(c)}$ is the central density of the
condensate, $\rho _{BE}^{(c)}=\rho _{BE}(0)$.

The mass profile $%
m_{BE}(r)=4\pi \int_{0}^{r}\rho _{BE}(r)r^{2}dr$ of the Bose-Einstein
condensate galactic halo is
\begin{equation}
m_{BE}\left( r\right) =\frac{4\pi \rho _{BE}^{(c)}}{k^{2}}r\left( \frac{\sin
kr}{kr}-\cos kr\right) ,
\end{equation}
with the boundary radius $R_{BE}$. At the boundary of the dark matter
distribution $\rho _{BE}(R_{BE})=0$, giving the condition $kR_{BE}=\pi $,
which fixes the radius of the condensate dark matter halo as
\be\label{rad}
R_{BE}=\pi
\sqrt{\frac{\hbar ^{2}a}{Gm_{\chi}^{3}}}.
\ee

The total mass $M_{BE}$ of the condensate is
\begin{equation}\label{mass1}
M_{BE}\left(R_{BE}\right)=4\pi \int_{0}^{R_{BE}}\rho _{BE}(r)r^{2}dr=\frac{4}{\pi }R_{BE}^{3}\rho _{BE}^{(c)},
\end{equation}
which represent a simple cubic proportionality between mass and radius. The mass of the condensate is around three times smaller than the mass of a constant  density sphere.

For $r>R_{BE}$ the density is smaller than zero, so that the $n=1$ polytropic density profile cannot be extended to infinity, that is, beyond its sharp boundary. For small values of $r$ the condensate dark matter profile can be written as
\be\label{app1}
\rho _{BE}(r)\approx \rho _{BE}^{(c)}\left(1 -
  \frac{{\pi }^2}{6}\frac{r^2}{R_{BE}^2}+ \frac{{\pi }^4}{120}\frac{r^4}{R_{BE}^4}+... \right), r\leq R_{BE}.
  \ee

  The gravitational potential $V^{(grav)}_{BE} \left( r\right) $ of the condensed dark
matter distribution is determined by the equation
\bea
V_{grav}^{(BE)} (r)&=&-G\int_{r}^{R_{BE}}\frac{m_{BE}\left( r^{\prime }\right) dr^{\prime }}{%
r^{\prime 2}}=\nonumber\\
&&-\frac{4G\rho _cR_{BE}^3}{\pi ^2r}
    \sin \left(\frac{\pi r}{R_{BE}}\right), r\leq R_{BE}.
\eea

At small radii the potential behaves as
\bea\label{app2}
V_{grav}^{(BE)} (r)&\approx& -G\rho _cR_{BE}^2\left(\frac{4}{\pi } -
  \frac{2\pi }{3}\frac{r^2}{R_{BE}^2} +
  \frac{\pi ^3}{30}\frac{r^4}{R_{BE}^4}\right) +\nonumber\\
 && O(r)^6, r\leq R_{BE}.
\eea

The tangential velocity of a test particle
moving in the condensed dark halo can be represented as \cite{BoHa07}
\be
V_{BE}^{2}\left(
r\right) =\frac{Gm_{BE}(r)}{r}= \frac{4\pi G\rho _{BE}^{(c)}}{k^{2}} \left(
\frac{\sin kr}{kr}-\cos kr\right).
\ee

With the use of Eq.~(\ref{rad}) the mass of the particle in the
condensate can be obtained from the radius of the dark matter halo in the
form \cite{BoHa07}
\begin{eqnarray}\label{mass}
m_{\chi}&=&\left( \frac{\pi ^{2}\hbar ^{2}a}{GR_{BE}^{2}}\right) ^{1/3}\approx \nonumber\\
&&2.58\times 10^{-30}\left[ a\left( \mathrm{cm}\right) \right] ^{1/3}\left[
R_{BE}\;\mathrm{(kpc)}\right] ^{-2/3}\mathrm{g}\approx
\nonumber\\
&& 6.73\times 10^{-2}%
\left[ a\left( \mathrm{fm}\right) \right] ^{1/3}\left[ R_{BE}\;\mathrm{(kpc)}%
\right] ^{-2/3}\mathrm{eV},
\end{eqnarray}
where $1\;{\rm g}=5.598\times 10^{32}\;{\rm eV}$.

From this equation it follows that $m_{\chi}$ is of the order of eV. For $a\approx 1
$ fm and $R_{BE}\approx 10$ kpc, the mass is of the order of $m_{\chi}\approx 14$
meV. For values of $a$ of the order of $a\approx 10^{6}$ fm, corresponding
to the values of $a$ observed in terrestrial laboratory experiments, $%
m_{\chi}\approx 1.44$ eV. These numerical values of the mass of the dark matter particle are perfectly consistent with the limit $%
m_{\chi}<1.87$ eV, obtained for the mass of the condensate particle from
cosmological considerations \cite{Bo}.

The properties of dark matter can be obtained observationally through  the study of the collisions between clusters of galaxies, like the Bullet Cluster (1E 0657-56) and the baby Bullet Cluster (MACSJ0025-12) \cite{Bul, Bul1}.
From these observations one can obtain constraints on the physical properties of the dark matter, such as its interaction cross-section with baryonic matter, and the dark matter-dark matter self-interaction cross section. If the ratio $\sigma _m=\sigma /m_{\chi}$ of the self-interaction cross section $\sigma =4\pi a^2$ and of the dark matter particle mass $m_{\chi }$ is known from observations, with the use of Eq.~(\ref{mass}) the mass of the dark matter particle in the Bose-Einstein condensate can be obtained  as \citep{Har5}
\begin{equation}
m_{\chi }=\left(\frac{\pi ^{3/2}\hbar ^2}{2G}\frac{\sqrt{\sigma _m}}{R_{BE}^2}\right)^{2/5}.
\end{equation}

The comparison of the results obtained from X-ray, strong lensing, weak lensing, and optical observations with numerical simulations
of the merging galaxy cluster 1E 0657-56, gives an upper limit (68 \% confidence) for $\sigma _m$ of the order of $\sigma _m<1.25\;{\rm cm^2/g}$ \cite{Bul}. By choosing  for $\sigma _m$ a value of $\sigma _m=1.25\;{\rm cm^2/g}$, we obtain for the mass of the dark matter particle an upper limit of the order
\begin{eqnarray}
m_{\chi }&<&3.1933\times10^{-37}\times\nonumber\\
&&\left(\frac{R_{BE}}{10\;{\rm kpc}}\right)^{-4/5}\times
\left(\frac{\sigma _m}{1.25\;{\rm cm^2/g}}\right)^{1/5}\;{\rm g}=\nonumber\\
&&0.1791\times\left(\frac{R_{BE}}{10\;{\rm kpc}}\right)^{-4/5} \times
\left(\frac{\sigma _m}{1.25\;{\rm cm^2/g}}\right)^{1/5}\;{\rm meV}.\nonumber\\
\end{eqnarray}
 By using this value of the particle mass we can estimate the scattering length $l_a$ as
\bea
a<\sqrt{\frac{\sigma _m\times m_{\chi }}{4\pi }}=1.7827\times 10^{-19}\;{\rm cm}=
1.7827\times 10^{-6}\;{\rm fm}.\nonumber\\
\eea

A stronger constraint for $\sigma _m$ was proposed in \cite{Bul1}, so that $\sigma _m\in(0.00335\;{\rm cm^2/g},0.0559\;{\rm cm^2/g})$, giving a dark matter particle mass of the order \cite{Har5}
\bea
m_{\chi }&\approx &\left(9.516\times 10^{-38}-1.670\times
10^{-37}\right)\times \nonumber\\
&&\left(\frac{R_{BE}}{10\;{\rm kpc}}\right)^{-4/5}\;{\rm g}=\nonumber\\
&& \left(0.053-0.093\right)\times
\left(\frac{R_{BE}}{10\;{\rm kpc}}\right)^{-4/5}\;{\rm meV},
\eea
and a scattering length of the order of
\bea
a&\approx &\left(5.038-27.255\right)\times 10^{-21}\;{\rm cm}=\nonumber\\
&&\left(5.038-27.255\right)\times 10^{-8}\;{\rm fm}.
\eea
Therefore the galactic radii data and the Bullet Cluster constraints predict a condensate dark particle mass of the order of $m_{\chi }\approx 0.1$ meV.

\section{Time-dependent evolution of zero temperature Bose-Einstein dark matter halos}\label{sect3}

In the present Section we introduce the variational formulation of the Gross-Pitaevskii equation, describing the time dynamics of the cold ($T=0$) Bose-Einstein condensed dark matter halos, and we obtain the basic equation describing the evolution of the gravitationally bounded dark matter condensates. We begin our analysis with the study of the general relativistic Bose-Einstein condensates trapped in a gravitational field in the semiclassical approximation, and with the use of the Newtonian limit we obtain the basic equations describing the gravitational collapse of the Bose-Einstein Condensates.

\subsection{The Newtonian limit for general relativistic Bose-Einstein Condensates}

In formulating a general relativistic model of the Bose-Einstein Condensates we consider that  bosonic matter at temperatures below the critical temperature $T_c$ in a gravitational field represents a hybrid system, in which the gravitational field
remains classical, while the bosonic condensate  is described by quantum fields.  In the standard approach used for coupling quantum fields to a classical gravitational field (i. e., semiclassical gravity), the energy momentum tensor that serves as the source in the
Einstein equations is replaced by the expectation value of the energy momentum operator
$\hat{T}_{\mu \nu}$ with respect to some quantum state $\Psi $ \cite{Carl},
\be\label{gr1}
R_{\mu \nu}-\frac{1}{2}g_{\mu \nu}R=\frac{8\pi G}{c^4}\left \langle\Psi \left |\hat{T}_{\mu \nu}\right |\Psi \right \rangle,
\ee
where $R_{\mu \nu}$ is the curvature tensor, $R$ is the curvature scalar, and $g_{\mu \nu}$ is the metric tensor of
the space-time. In the non-relativistic limit the state function $\Psi $
evolves according to the Gross-Pitaevski
equation for the condensate wave function $\psi $, with a quartic non-linear
term \cite{Da99,rev,Pit, Pet, Zar, BoHa07}
\bea \label{sch}
i\hbar \frac{\partial }{\partial t}\psi \left( \vec{r},t\right) &=&
\left[ -%
\frac{\hbar ^{2}}{2m_{\chi}}\nabla ^{2}+m_{\chi}V_{ext}(\vec{r},t)+u_{0}|\psi (\vec{r}%
,t)|^{2}\right]\times \nonumber\\
&& \psi (\vec{r},t),
\eea%
where $u_{0}=4\pi \hbar ^{2}a/m_{\chi}$, $a$ is the coherent scattering length
(defined as the zero-energy limit of the scattering amplitude $f_{scat}$), $%
m _{\chi}$ is the mass of the condensate particle, and $V_{ext}$ is the external
potential. For the energy-momentum tensor we obtain $\left \langle\Psi \left |\hat{T}_{\mu \nu}\right |\Psi \right \rangle=\left \langle\psi \left |\hat{T}_{\mu \nu}\right |\psi \right \rangle=T_{\mu \nu}$, where $T_{\mu \nu}$ is the classical energy-momentum tensor of the condensed system. In a comoving frame the energy momentum tensor is diagonal, with components $T_{\mu \nu}=\left(\rho c^2, -P,-P,-P\right)$, with $\rho c^2$ and $P$ denoting the effective energy density and the thermodynamic pressure of the condensed system. As for $V_{ext}(\vec{r},t)$, we assume that it is the
gravitational potential $V_{grav}$, $V_{ext}(\vec{r},t)=V_{grav}\left( \vec{r%
},t\right) $, and it is given by the Newtonian limit of Eq.~(\ref{gr1}). In this limit $g_{00}=1+2V_{grav}(\vec{r},t)/c^2$, and  Eq.~(\ref{gr1}) can be written as $R_0^0=\left(4\pi G/c^2\right)\rho $. By taking into account that $R_0^0=\left(1/c^2\right)\Delta V_{grav}(\vec{r},t)$, it follows that for a single component condensate the gravitational potential $V_{grav}\left( \vec{r},t\right) $ satisfies the Poisson equation
\be\label{Poi}
\nabla ^{2}V_{grav}(\vec{r}%
,t)=4\pi G\rho (\vec{r},t),
\ee
where
\be
\rho =m_{\chi}n\left( \vec{r}%
,t\right) =m_{\chi}\left\vert \psi \left( \vec{r},t\right) \right\vert ^{2},
\ee
 is the
mass density inside the Bose-Einstein condensate, $n\left( \vec{r},t\right) =\left\vert \psi \left( \vec{r},t\right) \right\vert ^{2}$
is the particle number density, and $G$ is the gravitational constant,
respectively. The probability density $\left\vert \psi \left( \vec{r},t\right) \right\vert ^{2}$ is normalized
according to $\int {n\left( \vec{r},t\right)d^{3}\vec{r}}=\int {\left\vert \psi \left( \vec{r},t\right) \right\vert ^{2}d^{3}\vec{r}}=N$, where $N$ is the total particle number in the condensate.

The Newtonian approximation is valid only if the condition $P<<\rho c^2$ is satisfied. The equation of state of the Bose-Einstein Condensate is $P(\rho )=U_{0}\rho ^{2}$, with $U_{0}=2\pi \hbar ^{2}a/m_{\chi }^{3}$ \cite{BoHa07}, which is a polytrope with index $n=1$. Therefore the general relativistic corrections can be ignored for condensate dark matter densities satisfying the constraint $U_0\rho <<c^2$, or, equivalently,
\bea\label{condNewt}
\rho &<<&\frac{c^2m_{\chi }^3}{2\pi \hbar ^2a}=1.289\times 10^{-15}\times \nonumber\\
&&\left(\frac{m_{\chi }}{10^{-32}\;{\rm g}}\right)^3\left(\frac{a}{10^{-7}\;{\rm cm}}\right)\;{\rm g/cm^3}.
\eea

The central densities of the galactic dark matter halos are of the order of $\rho _c\approx 10^{-27}\;{\rm g/cm^3}-10^{-24}\;{\rm g/cm ^3}$ \cite{Bul, Bul1}, and hence the Newtonian approximation can be applied for the study of the astrophysical properties of the condensate dark matter systems.

\subsection{Variational formulation of the Gross-Pitaevskii equation}

The basic equations describing the gravitationally trapped Bose-Einstein Condensates in the Newtonian limit are given by Eqs.~(\ref{sch}) and (\ref{Poi}), respectively.
The time dependent Gross-Pitaevski Eq. (\ref{sch}) can be derived from the
action principle $\delta \int_{t_{1}}^{t_{2}}L_{\psi}dt=0$, where the Lagrangian $L_{\psi}$
is given by \cite{Pet,var,Fetter}
\begin{equation}
L_{\psi}=\int _V{\frac{i\hbar }{2}\left( \psi ^{\ast }\frac{\partial \psi }{\partial t}%
-\psi \frac{\partial \psi ^{\ast }}{\partial t}\right) d^{3}\vec{r}}-E,  \label{lagr}
\end{equation}
where $E=\int _V\varepsilon d^{3}\vec{r}$ is the total energy, with the energy density $%
\varepsilon $ given by
\begin{equation}
\varepsilon =\frac{\hbar ^{2}}{2m_{\chi}}\left\vert \nabla \psi \right\vert
^{2}+m_{\chi}V_{grav}\left\vert \psi \right\vert ^{2}+\frac{u_{0}}{2}\left\vert
\psi \right\vert ^{4}.  \label{energy}
\end{equation}

If one multiplies Eq.~(\ref{sch}) by $\psi ^{\ast }$ and subtracts the
complex conjugate of the resulting equation, one arrives at the equation
\begin{equation}
\frac{\partial \left| \psi \right| ^{2}}{\partial t}+\nabla \cdot \left[
\frac{\hbar }{2im_{\chi}}\left( \psi ^{\ast }\nabla \psi -\psi \nabla \psi ^{\ast
}\right) \right] =0.  \label{cont0}
\end{equation}

Eq.~(\ref{cont0}) has the form of a continuity equation for the particle
density, and can be written as
\begin{equation}\label{18}
\frac{\partial \rho }{\partial t}+\nabla \cdot \left( \rho \vec{v}\right) =0,
\end{equation}
where the velocity of the condensate is defined by
\begin{equation}
\vec{v}=\frac{\hbar }{2im_{\chi}}\frac{\psi ^{\ast }\nabla \psi -\psi \nabla \psi
^{\ast }}{\left| \psi \right| ^{2}}.
\end{equation}

Simple expressions for the density and velocity can be obtained if we write
the wave function $\psi $ in terms of its amplitude $f$ and of a phase $%
\varphi $, $\psi =f\exp \left( i\varphi \right) $ (the Madelung
representation). In this representation $\rho =f^{2}$, and the velocity $%
\vec{v}$ is $\vec{v}=\left( \hbar /m_{\chi}\right) \nabla \varphi $. In the
Madelung representation of the wave function, the dynamics of the
Bose-Einstein condensate can be formulated in terms of the continuity
equation and of the hydrodynamic Euler equations \cite{BoHa07}.

\subsection{Dynamical time evolution of the Bose-Einstein condensed dark matter halos}

In order to study the time evolution of the condensate we apply the
variational approach, by assuming a trial form for the wave function $\psi $ of the condensate. We assume that during the motion of the dark matter cloud, the density profile
maintains its shape, but that its spatial extent $R=R(t)$ depends on time. The
density distribution is determined by the amplitude of the wave function,
while the phase determines its velocity field.

In the following we assume
that the velocity is in the radial direction, and it is proportional to $r$.
Translated into the behavior of the wave function, this implies that the
phase $\varphi $ varies as $r^{2}$, since the radial velocity of the
condensate is given by $\left( \hbar /m_{\chi}\right) \partial \varphi /\partial r$%
. We therefore write the phase of the wave function as $\varphi
=H(t)m_{\chi}r^{2}/2\hbar $, where $H(t)$ is a second parameter in the wave function,
which determines the velocity of the condensate as
\be\label{vel}
\vec{v}=H(t)\vec{r}.
\ee
Thus,
the complete trial wave function is \cite{Pet,var,Fetter}
\begin{equation}
\psi \left( r,t\right) =\frac{A\sqrt{N}}{R^{3/2}(t)}f\left[ \frac{r}{R(t)}\right]
\exp \left[ iH(t)\frac{m_{\chi}r^{2}}{2\hbar }\right] ,
\end{equation}
where $A$ is a normalization constant, and $f\left[r/R(t)\right]$ is an arbitrary
function to be determined. The mass density of the condensate is thus given by
\bea\label{rhotrial}
\rho (r,t)&=&m_{\chi}\left |\psi  \left( r,t\right)\right|^2=\frac{A^2Nm_{\chi}}{R^3(t)}f^2\left[ \frac{r}{R(t)}\right]=\nonumber\\
&&\frac{A^2M}{R^3(t)}f^2\left[ \frac{r}{R(t)}\right],
\eea
where $M=Nm_{\chi}$ is the total mass of the condensate dark matter halo.

We now carry out the integration over $r$ in Eq.~(\ref{lagr}) for
this choice of the wave function, and obtain the Lagrangian of the
condensate as a function of the two independent variables $H(t)$ and $R(t)$, as
well as of the time derivative of $H(t)$. For the first term in Eq.~(\ref{lagr}%
) we obtain
\bea
&&\int \frac{i\hbar }{2}\left( \psi ^{\ast }\frac{\partial \psi }{\partial t}%
-\psi \frac{\partial \psi ^{\ast }}{\partial t}\right) dV=\nonumber\\
&&-\frac{1}{2}\dot{H}(t)%
\int m_{\chi}n\left( \vec{r},t\right) r^{2}dV=\nonumber\\
&&-\frac{1}{2}m_{eff}R^{2}(t)\dot{H}(t),
\eea
where
\begin{equation}
m_{eff}(R)=Nm_{\chi}\frac{\left\langle r^{2}\right\rangle }{R^{2}(t)},
\end{equation}
and
\begin{equation}
\left\langle r^{2}\right\rangle =\frac{\int n\left( \vec{r},t\right) r^{2}d^3\vec{r}%
}{\int n\left( \vec{r},t\right) d^3\vec{r}},
\end{equation}
is the mean square radius of the dark matter halo. The total energy of the
condensate is obtained as
\bea
E&=&\frac{1}{2}m_{eff}(R)R^{2}(t)H^{2}(t)+\nonumber\\
&&E_{zp}(R)+E_{grav}(R)+E_{int}(R)=\nonumber\\
&&\frac{1}{2}%
m_{eff}(R)R^{2}(t)H^{2}(t)+U(R),
\eea
where
\be
U(R)=E_{zp}(R)+E_{grav}(R)+E_{int}(R),
\ee
and
\begin{equation}
E_{zp}(R)=\frac{\hbar ^{2}}{2m_{\chi}}\int _V {\left( \frac{d\left| \psi \right| }{dr}%
\right) ^{2}d^{3}\vec{r}}=C_{zp}R^{-2},  \label{c1}
\end{equation}
\begin{equation}
E_{grav}(R)=\int _V{ m_{\chi}V_{grav}\left\vert \psi \right\vert ^{2}d^{3}\vec{r}}=-C_{grav}R^{-1},
\label{c2}
\end{equation}%
and
\begin{equation}
E_{int}(R)=\frac{u_{0}}{2}\int _V{ \left| \psi \right| ^{4}d^{3}\vec{r}}=C_{int}R^{-3},
\label{c3}
\end{equation}
respectively. The coefficients $C_{zp}$, $C_{grav}$, and $C_{int}$,
respectively, are defined by  Eqs.~(\ref{c1})-(\ref{c3}), and are given by
\be\label{czp1}
C_{zp}=\frac{2\pi \hbar ^2A^2N}{m_{\chi}}\int_0^1{f^{\prime 2}(\xi )\xi ^2d\xi },
\ee
\be\label{cgrav1}
C_{grav}=4\pi m_{\chi} A^2N\int_0^1{f^2(\xi)V_{grav}(\xi )\xi ^2d\xi },
\ee
and
\be\label{cint1}
C_{int}=2\pi u_0A^4N^2\int_0^1{f^4(\xi)\xi ^2d\xi},
\ee
respectively.
 The
coefficient $C_{zp}$ describes the so-called zero point kinetic energy of the
system. Therefore the Lagrangian of the Bose-Einstein condensate dark matter
halo takes the form of an effective Lagrangian, which can be written as
\begin{equation}
L_{eff}\left(R,H,\dot{H}\right)=-\left[ \frac{1}{2}m_{eff}(R)R^{2}\left( \dot{H}+H^{2}\right) +U\left(
R\right) \right] .
\end{equation}

The Lagrange equation for $\dot{H}$,
\be
 \frac{d}{dt} \left[ \frac{\partial
L_{eff}\left(R,H,\dot{H}\right)}{\partial \dot{H}}\right] =\frac{\partial L_{eff}\left(R,H,\dot{H}\right)}{\partial H},
\ee
 gives
\begin{equation}
H=\frac{\dot{R}}{R},  \label{H}
\end{equation}
while the Lagrange equation for $R$, $\partial L_{eff}\left(R,H,\dot{H}\right)/\partial R=0$, gives the
equation of motion of the condensate as
\begin{equation}
 \left[\frac{1}{2}\frac{\partial m_{eff}(R)}{\partial R}R+m_{eff}\right] R\left( \dot{H}+H^{2}\right)=-\frac{\partial U\left( R\right) }{%
\partial R}.  \label{eqmot}
\end{equation}

By integrating Eq.~(\ref{18}) over the volume of the condensate, with the use of the Gauss theorem we obtain immediately the following particle conservation number equation,
\be\label{9}
\frac{\partial }{\partial t}N+\int_S{\frac{\rho }{m_{\chi}}\vec{v}\cdot \vec{n}dS}=0,
\ee
where the total particle number  $N=\int _V{\left\vert \psi \left( \vec{r},t\right) \right\vert ^{2}d^{3}\vec{r}}$ is obtained by integrating the norm of the wave function over the entire spherical volume $V$ of the condensate, with radius $R_{BE}$. The time variation of $N$ can be due only to the gain or loss of the particles moving in or out of the condensate through the surface $S$ encompassing the volume $V$. In the case of a static condensate, $\vec{v}\equiv 0$, and $\rho \left(R_{BE}\right)\equiv 0$, and then obviously the particle flux $\vec{j}\left(R_{BE}\right)=\rho \left(R_{BE}\right)\vec{v}\equiv 0$, leading, via Eq.~(\ref{9}),  to the conservation of the total particle number in the system, $N={\rm constant}$.

Based on the static case, for a time evolving  condensate, we would expect that on its boundary the condition $\rho \left[R_{BE}(t)\right]\equiv 0$ to be satisfied for all times, leading to a zero particle flux through the surface encompassing the time dependent volume $V$ of the condensate, $\vec{j}\left[R_{BE}(t)\right]=\rho \left[R_{BE}(t)\right]\vec{v}[R_{BE}(t)]\equiv 0, \forall t\geq 0$.   Since the Hamiltonian of the Gross-Pitaevskii equation Eq.~(\ref{sch}) is real, it is obviously Hermitian, and therefore the norm of the wave function, as well as the total particle number, are conserved. Since $N$ is a constant, from Eq.~(\ref{9}) it follows that in the time dependent case the particle flux through the boundary $S$ of the condensate must be zero, that is, $\vec{j}\left[R_{BE}(t)\right]=\rho \left[R_{BE}(t)\right]\vec{v}\left[R_{BE}(t)\right]\equiv 0, \forall t\geq 0$, and there is no particle loss from the system.

\section{Dark matter density, gravitational potential and physical parameters of time-dependent Bose-Einstein Condensate dark matter halos}\label{sect4}

 As a first step in the study of the time dynamics of the
gravitationally bounded Bose - Einstein condensates we have to chose a
variational trial wave function. Instead of fixing it in an arbitrary way
(by assuming, for example, that the initial density profile of the
condensate has a Gaussian form), we require that $\left| \psi \right| ^{2}$
satisfies the continuity equation Eq.~(\ref{18}). For the density of the
Bose-Einstein condensate we assume a general form
\begin{equation}\label{dens1}
\rho \left( r,t\right) =\rho _{0}\left( t\right) +\rho _{1}(t)\rho _{2}(r),
\end{equation}
where $\rho _{0}(t)$, $\rho _{1}(t)$ and $\rho _{2}(r)$ are arbitrary
functions of $t$ and $r$ to be determined. From a physical point of view, the trial density profile $\rho (t)$ is the sum of two terms, the first representing a "cosmological" type homogenous term $\rho _{hom}(t)$, while  the second term $\rho _{inhom}(t,r)$ represents the effect of the time-dependent inhomogeneities in the dark matter halo. The inhomogeneous term is assumed to be separable in the variables $t$ and $r$, so that $\rho _{inhom}(t,r)=\rho _1(t)\rho _2(r)$.

\subsection{The density profile and the gravitational potential of the dark matter halo}

Substitution of Eqs.~(\ref{vel}) and (\ref{dens1}) into the
continuity equation Eq.~(\ref{18}) gives
\bea
&&\dot{\rho}_{0}(t)+3\rho _{0}(t)\frac{\dot{R}(t)}{R(t)}+\rho _{2}(r)\times \nonumber\\
&&\left[
\dot{\rho}_{1}(t)+3\rho _{1}(t)\frac{\dot{R}(t)}{R(t)}+\frac{\dot{R}(t)}{R(t)%
}\rho _{1}(t)r\frac{\rho _{2}^{\prime }(r)}{\rho _{2}(r)}\right] =0.
\label{cont}
\eea

We determine the function $\rho _{2}(r)$ by imposing the condition $r\rho
_{2}^{\prime }(r)/\rho _{2}(r)={\rm constant}=\alpha >0$, which leads first to
\begin{equation}
\rho _{2}(r)=C_{1}r^{\alpha },
\end{equation}
where $C_{1}$ is an arbitrary constant of integration. Next we require that
the term in the square bracket of Eq.~(\ref{cont}) vanishes. Therefore, Eq.~(\ref{cont}) gives the following two independent differential equations for the determination of the functions $\rho _0(t)$ and $\rho _1(t)$,
\be
\dot{\rho}_{0}(t)+3\rho _{0}(t)\frac{\dot{R}(t)}{R(t)}=0,
\ee
and
\be
\dot{\rho}_{1}(t)+\left(3+\alpha \right)\rho _{1}(t)\frac{\dot{R}(t)}{R(t)}=0,
\ee
respectively.
Hence, the
general solution of Eq.~(\ref{cont}) can be obtained as
\begin{equation}
\rho \left( r,t\right) =\frac{1}{R^{3}(t)}\left[ a_{0}+b_{0}\frac{r^{\alpha }%
}{R^{\alpha }(t)}\right] ,
\end{equation}
where $a_{0}$ and $b_{0}$ are arbitrary constants of integration. Since at
the vacuum boundary of the condensate, where $r=R(t)$, the density
must satisfy the condition $\rho \left[ R(t),t\right] =0$, $\forall t\geq 0$, we obtain for the two integration constants the
condition $a_{0}+b_{0}=0$. Therefore the density profile of the condensate
can be represented in the form
\begin{equation}
\rho \left( r,t\right) =\frac{a_{0}}{R^{3}(t)}\left[ 1-\frac{r^{\alpha }}{%
R^{\alpha }(t)}\right] .
\end{equation}
With this density profile the Poisson equation can be written as
\begin{equation}
\frac{1}{r^{2}}\frac{d}{dr}\left( r^{2}\frac{dV_{grav}}{dr}\right) =4\pi
Ga_{0}\frac{1}{R^{3}(t)}\left[ 1-\frac{r^{\alpha }}{R^{\alpha }(t)}\right] ,
\end{equation}
and can be integrated to give
\begin{equation}
\frac{dV_{grav}}{dr}=4\pi Ga_{0}\frac{1}{R^{3}(t)}\left[ \frac{r}{3}-\frac{%
r^{\alpha +1}}{\left( \alpha +3\right) R^{\alpha }(t)}\right] +\frac{C_{2}(t)%
}{r^{2}},
\end{equation}
where $C_2(t)$ is an arbitrary integration function.
In order to avoid any singularities for $r\rightarrow 0$ we take $%
C_{2}(t)\equiv 0$. A new integration gives
\bea
V_{grav}\left( r,t\right) &=&4\pi Ga_{0}\frac{1}{R^{3}(t)}\times \nonumber\\
&&\left[ \frac{r^{2}}{6%
}-\frac{r^{\alpha +2}}{\left( \alpha +2\right) \left( \alpha +3\right)
R^{\alpha }(t)}\right] +
V_{grav}^{(0)}(t). \nonumber\\
\eea

At the surface of the dark matter halo $r=R(t)$ the gravitational potential
becomes $V_{grav}\left[ R(t),t\right] =-GM/R(t)$, where $M=Nm_{\chi}$ is the total
mass of the condensate, which gives
\begin{equation}
V_{grav}^{(0)}(t)=-\frac{GM+2\pi Ga_{0}\left[ \alpha \left( \alpha +5\right)
/3\left( \alpha +2\right) \left( \alpha +3\right) \right] }{R(t)},
\end{equation}
leading to
\bea
V_{grav}\left( r,t\right)& =&\frac{\pi Ga_{0}}{R^{3}(t)}\left[ \frac{2r^{2}}{3%
}-\frac{4r^{\alpha +2}}{\left( \alpha +2\right) \left( \alpha +3\right)
R^{\alpha }(t)}\right] -\nonumber\\
&&\frac{GM+2\pi Ga_{0}\left[ \alpha \left( \alpha
+5\right) /3\left( \alpha +2\right) \left( \alpha +3\right) \right] }{R(t)}.\nonumber\\
\eea

Since the integration constant $a_{0}$ is arbitrary, we determine it by
imposing the condition
\be
\pi Ga_{0}=GM+4\pi Ga_{0}\left[ \alpha \left( \alpha
+5\right) /6\left( \alpha +2\right) \left( \alpha +3\right) \right] ,
\ee
giving
\begin{equation}
a_{0}=\frac{3\left( \alpha +2\right) \left( \alpha +3\right) }{\pi \left(
\alpha ^{2}+5\alpha +18\right) }M.  \label{a01}
\end{equation}

From the normalization condition $M=4\pi \int_{0}^{R}\rho r^{2}dr$ we obtain
\begin{equation}
a_{0}=\frac{3}{4\pi }\frac{\alpha +3}{\alpha }M.  \label{a02}
\end{equation}

By equating Eqs.~(\ref{a01}) and (\ref{a02}) we find for $\alpha $ the
value $\alpha =2$, giving $a_0=15M/8\pi $. Therefore the density and the gravitational potential in
the Bose-Einstein condensed dark matter halo become
\begin{equation}\label{rhocont}
\rho \left( r,t\right) =\frac{15}{8\pi }\frac{M}{R^{3}(t)}\left[ 1-\frac{%
r^{2}}{R^{2}(t)}\right] ,
\end{equation}
and
\begin{equation}
V_{grav}\left( r,t\right) =-\frac{15}{8}\frac{GM}{R}\left[ 1-\frac{2}{3}%
\frac{r^{2}}{R^{2}(t)}+\frac{1}{5}\frac{r^{4}}{R^{4}(t)}\right] ,
\end{equation}%
respectively. For the wave function of the condensate we obtain
\begin{equation}
\left\vert \psi \left( r,t\right) \right\vert ^{2}=\frac{15}{8\pi }\frac{N}{%
R^{3}}\left[ 1-\frac{r^{2}}{R^{2}(t)}\right] .
\end{equation}

By comparing Eqs.~(\ref{rhotrial}) and (\ref{rhocont}) we obtain the expressions of the trial function $f\left[r/R(t)\right]$ and of the normalization constant $A$ as
\be\label{trial}
A=\sqrt{\frac{15}{8\pi}},f\left[\frac{r}{R(t)}\right]=\sqrt{1-\frac{r^2}{R^2(t)}}.
\ee

\subsection{The coefficients $C_{zp}$, $C_{grav}$ and $C_{int}$}

Once the wave function of the system is known, the effective mass $m_{eff}$
and the constant coefficients in Eqs.~(\ref{czp1})-(\ref{cint1}) can be obtained
immediately. For the effective mass $m_{eff}$ we find
\begin{equation}
m_{eff}=\frac{3}{7}M={\rm constant}.
\end{equation}

The coefficient $C_{zp}$, giving the zero point kinetic energy of the condensate is given by
\be
C_{zp}=\frac{15\hbar ^{2}}{4m_{\chi}}\frac{M}{m_{\chi}}F\left( \xi \right),
\ee
where
\be
F\left( \xi \right) =\int_{0}^{\xi }{\frac{x^{4}}{\left(
1-x^{2}\right)}dx }.
\ee

In obtaining the zero point kinetic energy $E_{zp}$ by using the trial
function given by Eq.~(\ref{trial}), it turns out that the coefficient $C_{zp}
$ diverges at the surface of the condensate, where $r\rightarrow R(t)$. The
divergence of the kinetic energy near the edge of the condensate cloud is a
general feature of the Bose-Einstein condensates, and it can be understood
from simple physical considerations \cite{Pet}. In the Thomas-Fermi approximation the
particle number profile is given by $n(\vec{r})=\left[ \mu -V_{grav}(\vec{r})%
\right] /u_{0}$, where $\mu $ is the chemical potential of the condensate.
If $\vec{r}_{0}$ is a point on the surface, then $\mu =V_{grav}(\vec{r}_{0})$
defines the surface of the condensate. By expanding the external potential
about $\vec{r}_{0}$ we obtain $n\left( \vec{r}\right) =\vec{F}_{grav}\cdot \left(
\vec{r}-\vec{r}_{0}\right) /u_{0}$, where $\vec{F}_{grav}=-\nabla V_{grav}\left(
\vec{r}_{0}\right) $, while the wave function is given by $\psi \left(
r\right) =\left[ \vec{F}_{grav}\cdot \left( \vec{r}-\vec{r}_{0}\right) /u_{0}\right]
^{1/2}$ \cite{Pet,Pet1}. If we denote the coordinate in the direction of $\nabla
V\left( \vec{r}_{0}\right) $ by $x$, and denote the position of the surface
as $x_{0}$, the interior of the condensed dark matter halo corresponds to $%
x\leq x_{0}$. The wave function near the surface varies as $\left(
x_{0}-x\right) ^{1/2}$, and hence the kinetic energy
term in the energy functional behaves as $\hbar
^{2}\left\vert d\psi /dx\right\vert ^{2}/2m_{\chi }\left\vert \psi
\right\vert ^{2}\sim \hbar ^{2}/2m_{\chi }\left( x_{0}-x\right) ^{2}$, and
it diverges as $x$ approaches $x_{0}$ from below. In fact the kinetic energy
term dominates over the potential energy term for $x_{0}-x\leq \delta $,
where
\be\label{cutoff}
\delta =\left(\frac{ \hbar ^{2}}{2m_{\chi }F_{grav}}\right) ^{1/3}.
\ee

 In order to
obtain a better description of the surface properties of the Bose-Einstein Condensates one must study
the Gross-Pitaevskii equation \cite{Pet,Pet1}
\be
-\left( \frac{\hbar ^2}{2m_{\chi}}\right) \psi
^{\prime \prime }(x)+F_{grav}x\psi +u_{0}\left\vert \psi \right\vert ^{2}\psi =0,
\ee
describing the surface properties of the condensate.
An approximate solution of this equation is $\Psi =\sqrt{-y}$, $y\leq 0$,
where $y=x/\delta $, and $\Psi =\psi /\sqrt{F_{grav}\delta /u_{0}}$. By using this
solution one can evaluate the kinetic energy per unit area perpendicular to
the $x$-axis, $\left\langle p^{2}/2m_{\chi }\right\rangle =\left( \hbar
^{2}/2m_{\chi }\right) \int \left\vert \bigtriangledown \psi \right\vert
^{2}dx$. Since this integral diverges for $x\rightarrow 0$, the integral
must be evaluated for $x$ less than some cutoff value $-l$, with the lower limit
of integration taken as $-L$, where $L>>\delta $. By choosing the cut-off
distance as $-\delta $, with the use of the approximate solution for $\psi $
given above we obtain  $\left\langle p^{2}/2m_{\chi }\right\rangle \approx
\left( \hbar ^{2}/8m_{\chi }\right) \left( F_{grav}/u_{0}\right) \ln \left(
L/l\right) $. If one uses the true wave function obtained numerically one
finds $\left\langle p^{2}/2m_{\chi }\right\rangle \approx \left( \hbar
^{2}/8m_{\chi }\right) \left( F_{grav}/u_{0}\right) \ln \left( 4.160L/l\right) $ \cite{Pet}.
Hence in order to calculate the kinetic energy in more general situations an
effective cut-off must be used, whose numerical value is of the order of $l=0.240\delta$ \cite{Pet, Pet1}.

In order to make the integral $F(\xi)$ convergent, $\xi $ must
be smaller than $1$, $\xi <1$, and therefore we must introduce a cut-off length $\epsilon $, so that the upper integration limit becomes $\xi =1-\epsilon$. Since the force acting on a particle on the surface of the condensate is the gravitational force $F=GMm_{\chi}/R^2$, by taking into account the previous results we obtain the cut-off length $\epsilon =l/R$ as
\bea
\epsilon &=&0.240\frac{\delta }{R}=0.190488\times \left( \frac{\hbar ^2}{G m_{\chi}^2 M R}\right)^{1/3}=\nonumber\\
&&2.653\times 10^{-16}\times \left(\frac{m_{\chi}}{10^{-32}\;{\rm g}}\right)^{-2/3}\left(\frac{M}{10^6M_{\odot}}\right)^{-1/3}\times \nonumber\\
&&\left(\frac{R}{10\;{\rm kpc}}\right)^{-1/3}.
\eea

For $\epsilon =2.653\times 10^{-16}$ we obtain $F\left(1-\epsilon\right)=17.0350669$, while for $\epsilon =10^{-10}$ we have $F\left(1-\epsilon\right)=10.5261657$. By assuming that the surface region of the dark matter halo represents  around 1\% of its spatial extension, we have $\epsilon =0.01$, and $F\left( 1-\epsilon \right) =1.333$. In the following we consider two limiting cases for $\epsilon$, $\epsilon =0.01$ and $\epsilon =2.653\times 10^{-16}$, respectively.

Therefore the coefficients $C_{zp}$, $C_{grav}$ and $C_{int}$ can be obtained as
\bea
C_{zp}&=&\frac{15\hbar ^{2}}{4m_{\chi}}\frac{M}{m_{\chi}}F\left( \xi \right) =\left(1.110\times
10^{50}-1.419\times 10^{51}\right)\times \nonumber\\
&&\left( \frac{m_{\chi }}{10^{-32}\;{\rm g}}\right) ^{-2}\left( \frac{M}{%
10^{6}M_{\odot}}\right) {\rm \;g\;cm^{4}}/s^{2},  \label{c4}
\eea
\begin{equation}
C_{grav}=\frac{10}{7 }GM^{2}=3.81\times 10^{71}\times \left( \frac{M}{%
10^{6}M_{\odot}}\right) ^{2}{\rm \;g\;cm^{3}}/s^{2},  \label{c5}
\end{equation}%
and
\bea \label{c6}
C_{int}&=&\frac{15}{4\pi}u_{0}\frac{M^{2}}{m_{\chi }^{2}}=
6.665\times 10^{114}\times \nonumber\\
&&\left( \frac{a}{10^{-7}\;{\rm cm}}\right)
\left( \frac{m_{\chi }}{10^{-32}\;{\rm g}}\right) ^{-3}\left( \frac{M}{10^{6}M_{\odot}}\right)
^{2}{\rm \;g\;cm^{5}}/s^{2}, \nonumber\\
\eea
respectively.

\section{Gravitational collapse of Bose-Einstein condensate dark matter halos}\label{sect5}

By taking into account the explicit form of $H$, given by Eq.~(\ref{H}), and the constancy of the effective mass $m_{eff}$, the
equation of motion of the gravitationally bounded Bose-Einstein condensate
is given by
\begin{equation}
m_{eff}\ddot{R}=\frac{2C_{zp}}{R^{3}}-\frac{C_{grav}}{R^{2}}+\frac{3C_{int}}{%
R^{4}},  \label{mot1}
\end{equation}

Eq.~(\ref{mot1}) can be rewritten in the form of an energy conservation
equation as
\begin{equation}
\frac{1}{2}m_{eff}\dot{R}^{2}(t)+U\left( R\right) =E_{0}={\rm constant},  \label{cons}
\end{equation}
where
\begin{equation}
U(R)=\frac{C_{zp}}{R^{2}}-\frac{C_{grav}}{R}+\frac{C_{int}}{R^{3}}.
\end{equation}

\begin{figure}[!ht]
\includegraphics[width=0.98\linewidth]{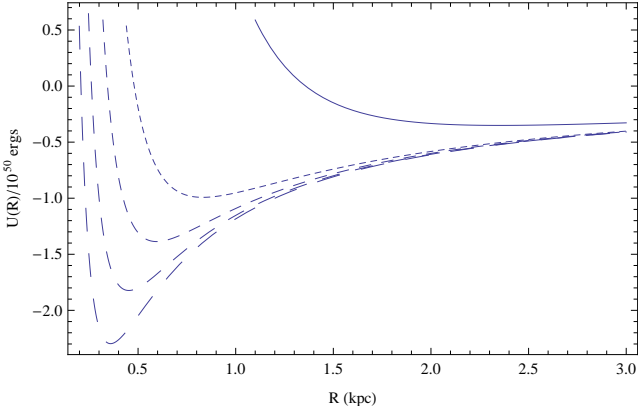}
\caption{Variation of the potential $U(R)$ as a function of $R$ for a Bose Einstein condensate dark matter halo with mass $M=10^6M_{\odot}$ and scattering length $a=10^{-7}$ cm. The mass of the condensate dark matter particle is $m= 10^{-32}$ g (solid curve), $m=2\times 10^{-32}$ g (dotted curve), $m=2.5\times 10^{-32}$ g (short dashed curve),  $m=3\times 10^{-32}$ g (long dashed curve), and $m=3.5\times 10^{-32}$ g (ultra long dashed curve), respectively.} \label{fig1}
\end{figure}

The variation of the potential $U(R)$ as a function of $R$ is represented in
Fig.~\ref{fig1}.

Let us consider the case of the contraction, that is, the case
of a Bose-Einstein dark matter halo beginning to contract at $t=0$ from an
initial radius $R=R_{0}=R_{\max }$. Then the initial velocity $\dot{R}(0)$ of the dark matter halo can be obtained from Eq.~(\ref{cons}) as
\be
\dot{R}(0)=\sqrt{\frac{2}{m_{eff}}\left[E_0-U\left(R_0\right)\right]}.
\ee

After a time $t_{coll}$ the condensate will eventually reach a stable radius $R_{st}$, where its velocity is zero, and the
potential energy $U(R)$ is minimum, $\left.\partial U/\partial R\right|_{R=R_{st}}=0$,
so that
\begin{equation}
\left. \frac{2C_{zp}}{R^{3}}-\frac{C_{grav}}{R^{2}}+\frac{3C_{int}}{R^{4}}%
\right|_{R=R_{st}}=0.  \label{eq}
\end{equation}

From Eq.~(\ref{eq}) we obtain the radius of the stable configuration as
\bea
R_{st}&=&\frac{C_{zp}+ \sqrt{C_{zp}^{2}+3C_{grav}C_{int}}}{C_{grav}}=\nonumber\\
&&\frac{%
C_{zp}}{C_{grav}}\left[ 1+ \sqrt{1+\frac{3C_{grav}C_{int}}{C_{zp}^{2}}}\right] .
\label{radius}
\eea

Since $3C_{grav}C_{int}/C_{zp}^2>>1$, the equilibrium radius can be obtained, in a very good approximation,
as
\bea\label{Rst}
R_{st}&\approx& \sqrt{\frac{3C_{int}}{C_{grav}}}=\sqrt{\frac{63}{8\pi }\frac{%
u_{0}}{Gm_{\chi}^{2}}}=\sqrt{\frac{63}{2}\frac{\hbar ^{2}a}{Gm_{\chi}^{3}}}=\nonumber\\
&&1.106\times
10^{21}\left( \frac{a}{10^{-7}\;{\rm cm}}\right) \left( \frac{m_{\chi}}{10^{-32}\;{\rm g}}\right)
^{-3}\;{\rm cm}.\nonumber\\
\eea

The condition of the zero final velocity of the halo, $\dot{R}\left(t_{coll}\right)=0$ determines the total energy
 $E_{0}$ as given by
\begin{equation}
E_{0}=U\left( R_{st}\right) =\frac{C_{zp}}{R_{st}^{2}}-\frac{C_{grav}}{R_{st}}+%
\frac{C_{int}}{R_{st}^{3}}.
\end{equation}

Therefore the initial velocity of the halo $\dot{R}(0)$ is determined as a function of the final radius $R_{st}$ and of the initial radius $R_0$ by the consistency condition (the conservation of energy),
\bea
\frac{m_{eff}}{2}\dot{R}^2(0)&=&\frac{C_{zp}}{R_{st}^{2}}-\frac{C_{grav}}{R_{st}}+%
\frac{C_{int}}{R_{st}^{3}}-\nonumber\\
&&\frac{C_{zp}}{R_{0}^{2}}+\frac{C_{grav}}{R_{0}}-%
\frac{C_{int}}{R_{0}^{3}}.
\eea

\begin{figure}[h]
\includegraphics[width=0.98\linewidth]{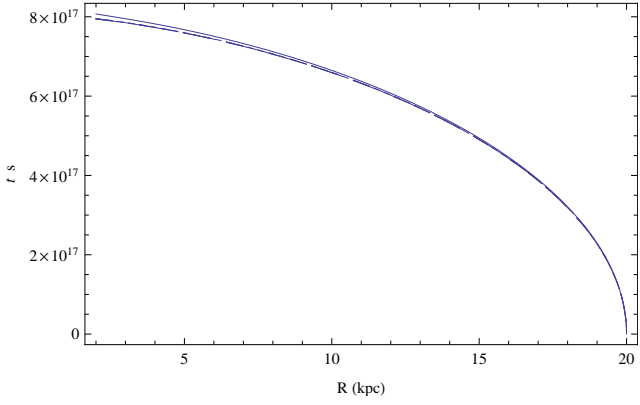}
\caption{Collapse time $t$ as a function of $R$ for a Bose-Einstein condensate dark matter halo with mass $M=10^6M_{\odot}$. The mass of the condensate dark matter particle is in the range $m_{\chi}\in \left(10^{-32}\;{\rm g}, 3.5\times 10^{-32}\;{\rm g}\right)$, and the scattering length is taken as $a=10^{-7}$ cm.} \label{fig2}
\end{figure}

The general solution of Eq.~(\ref{cons}) is given by
\begin{equation}
t-t_{0}=\pm \sqrt{\frac{m_{eff}}{2}}\int_{R_{0}}^{R}\frac{dR}{\sqrt{%
E_{0}-U\left( R\right) }},
\end{equation}
where $+$ and $-$ corresponds to the case of expansion and contraction,
respectively. The variation of the collapse time as a function of radius is presented, for $R_0=20$ kpc and $t_0=0$,  in Fig.~\ref{fig2}.

Since $E_0>U(R)$, $\forall R<R_0$, the function $1/\sqrt{E_0-U(R)}=1/\sqrt{E_0\left[1-U(R)/E_0\right]}$ can be approximated in the first order as
\bea
&&\frac{1}{\sqrt{E_0-U(R)}}\approx \frac{1}{\sqrt{E_0}}\left[1+\frac{U(R)}{2E_0}\right]=\nonumber\\
&&\frac{1}{\sqrt{E_0}}\left(1+\frac{C_{zp}}{2E_0}\frac{1}{R^2}-\frac{C_{grav}}{2E_0}\frac{1}{R}+\frac{C_{int}}{2E_0}\frac{1}{R^3}\right).
\eea

Therefore, within the framework of this approximation, the total collapse time
\be
t_{coll}=t\left(R_{st}\right)-t_0=\sqrt{\frac{m_{eff}}{2}}\int_{R_{0}}^{R_{st}}\frac{dR}{\sqrt{%
E_{0}-U\left( R\right) }},
\ee
 of the condensed dark matter halo between an initial radius $R_0$ and the radius of the stable configuration $R_{st}$ is given by
\begin{widetext}
\be
t_{coll}\approx \frac{\sqrt{m_{eff}} \left\{2 C_{grav}^{(0)} R_0^2 R_{st}^2 \ln \left(R_{st}/R_0\right)+\left(R_0-R_{st}\right) \left[C_{int}^{(0)} \left(R_0+R_{st}\right)+2
   R_0 R_{st} \left(C_{zp}^{(0)}+R_0 R_{st}\right)\right]\right\}}{2 \sqrt{2} \sqrt{E_0} R_0^2 R_{st}^2},
\ee
\end{widetext}
where we have denoted $C_{zp}^{(0)}=C_{zp}/2E_0$, $C_{grav}^{(0)}=C_{grav}/2E_0$, and $C_{int}^{(0)}=C_{int}/2E_0$, respectively. By using Eq.~(\ref{Rst}) to express the radius of the final stable configuration, we obtain for the collapse time the expression
\bea
t_{coll}&\approx& \frac{\sqrt{m_{eff}}}{6\sqrt{2}C_{int}^{(0)}\sqrt{E_{0}}%
R_{0}^{2}}\times
\Bigg\{ 2\sqrt{3}C_{zp}^{(0)}R_{0}^{2}\sqrt{%
C_{grav}^{(0)}C_{int}^{(0)}}+\nonumber\\
&&C_{int}^{(0)}R_{0}\Bigg[ R_{0}\left(
C_{grav}^{(0)}+6R_{0}-6\sqrt{3}\sqrt{\frac{C_{int}^{(0)}}{C_{grav}^{(0)}}}%
\right) -\nonumber\\
&&6C_{zp}^{(0)}\Bigg] +
3C_{grav}^{(0)}C_{int}^{(0)}R_{0}^{2}\ln
\left( \frac{3C_{int}^{(0)}}{C_{grav}^{(0)}R_{0}^{2}}\right)
-\nonumber\\
&&3(C_{int}^{(0)})^{2}\Bigg\} .
   \eea

   If the initial size of the condensed dark matter cloud $R_0$ is very high, so that the condition $E_0=C_{zp}/R_0^2-C_{grav}/R_0+C_{int}/R_0^3\approx 0$ is satisfied, and by assuming $C_{zp}=0$, the collapse time of the condensate can be expressed in an exact form as
   \begin{widetext}
   \bea
t_{coll}&\approx &\frac{\sqrt{2m_{eff}}}{3C_{grav}^{3}}\times \nonumber\\
&&\Bigg\{ \frac{\sqrt{%
R_{st}}\left[ \left( C_{grav}^{3}C_{int}\right) ^{3/4}\sqrt{R_{st}-%
\frac{C_{int}}{C_{grav}R_{st}}}F\left( \left. \sin ^{-1}\left( \frac{\sqrt[4]%
{\frac{C_{int}}{C_{grav}}}}{\sqrt{R_{st}}}\right) \right\vert -1\right)
+C_{grav}^{2}\left( C_{int}-C_{grav}^{2}R_{st}\right) \right] }{\sqrt{%
C_{grav}^{2}R_{st}^{2}-C_{int}}}-\nonumber\\
&&\frac{\sqrt{R_{0}}\left[ \left(
C_{grav}^{3}C_{int}\right) ^{3/4}\sqrt{R_{0}-\frac{C_{int}}{C_{grav}R_{0}}}%
F\left( \left. \sin ^{-1}\left( \frac{\sqrt[4]{\frac{C_{int}}{C_{grav}}}}{%
\sqrt{R_{0}}}\right) \right\vert -1\right) +C_{grav}^{2}\left(
C_{int}-C_{grav}^{2}R_{0}\right) \right] }{\sqrt{%
C_{grav}^{2}R_{0}^{2}-C_{int}}}\Bigg\},
\eea
\end{widetext}
where $F(\phi|m)$ denotes the elliptic integral of the first kind, $F(\phi|m)=\int_0^{\phi }{\left(1-m\sin ^2\theta\right)^{-1/2}d\theta }$. By using Eq.~(\ref{Rst}) for the equilibrium radius, the collapse time can be written as
\begin{widetext}
\bea
t_{coll}&\approx&  \frac{\sqrt{2m_{eff}}C_{int}^{3/4}}{3C_{grav}^{5/4}}\Bigg\{
C_{grav}\sqrt{\frac{R_{0}}{C_{int}^{3/2}}}\sqrt{C_{grav}R_{0}^{2}-C_{int}}-%
C_{int}^{5/4}\left[ \sqrt{2}\sqrt[4]{3}-F\left( \left. \csc ^{-1}\left( \sqrt[4]{3}%
\right) \right\vert -1\right) \right] -\nonumber\\
&&C_{int}^{5/4}F\left( \left. \sin ^{-1}\left( \frac{%
\sqrt[4]{\frac{C_{int}}{C_{grav}}}}{\sqrt{R_{0}}}\right) \right\vert
-1\right) \Bigg\} .
\eea
\end{widetext}

The numerical values of the elliptic integral of the first kind are of the order of unity. By neglecting the numerical factors of the order of unity, and by considering that the initial radius $R_0$ of the dark matter halo is enough high so that the condition $C_{grav}R_0^2>>C_{int}$ is satisfied, the collapse time can be obtained in the simple form of
\be
t_{coll}\approx \frac{\sqrt{2m_{eff}}}{3}C_{grav}^{-1/2}R_0^{3/2},E_0\approx 0, R_0>>\sqrt{\frac{C_{int}}{C_{grav}}}.
\ee
With the use of the expression of $C_{grav}$ we obtain
\bea
t_{coll}&\approx& \frac{1}{3}\sqrt{\frac{3}{5}}\frac{R_0^{3/2}}{\sqrt{GM}}=3.821\times 10^{18}\times \nonumber\\
&&\left(\frac{R_0}{100\;{\rm kpc}}\right)^{3/2}\times \left(\frac{M}{10^6M_{\odot}}\right)^{-1/2}\;{\rm s}, \nonumber\\
&&E_0\approx 0, R_0>>\left( \frac{a}{10^{-7}\;{\rm cm}}\right) \left( \frac{m_{\chi}}{10^{-32}\;{\rm g}}\right)
^{-3}\;{\rm kpc}.\nonumber\\
\eea
In this approximation the collapse time of the dark matter halo is determined only by the gravitational properties of the system, and it is independent of the zero point kinetic energy and the interaction energy.

The total mass $M_{st}$ of the stable condensate configuration with radius $R_{st}$, at the end of the collapse process, can be approximated as
\bea
&&M_{st}=\frac{4}{\pi}R_{st}^3\rho _c\approx \frac{108}{\pi}\left(\frac{7}{2}\right)^{3/2}\frac{\hbar^3}{m^{9/2}_{\chi}}\left(\frac{a}{G}\right)^{3/2}\rho _c=\nonumber\\
&&2.419\times 10^8\times
\left(\frac{m_{\chi}}{10^{-32}\;{\rm g}}\right)^{-9/2}\left(\frac{a}{10^{-7}\;{\rm cm}}\right)^{3/2}\times\nonumber\\
&&\left(\frac{\rho _c}{10^{-24}\;{\rm g/cm}^3}\right)M_{\odot},
\eea
where $\rho _c$ is the central density of the dark matter halo.

At the equilibrium point $R=R_{st}$ we can easily obtain the relations
\be\label{eqosc1}
\left.R\frac{dU}{dR}\right|_{R=R_{st}}=2E_{zp}\left(R_{st}\right)+E_{grav}\left(R_{st}\right)+3E_{int}\left(R_{st}\right)=0,
\ee
and
\be\label{eqosc2}
\left.R^2\frac{d^2U}{dR^2}\right|_{R=R_{st}}=6E_{zp}\left(R_{st}\right)+2E_{grav}\left(R_{st}\right)+12E_{int}\left(R_{st}\right),
\ee
respectively.

\section{The stability of the  collapsed Bose-Einstein Condensate dark matter halos}\label{sectnn}

For condensed dark matter halos with masses of the order of $M=10^8M_{\odot}$ and radii of the order of $R=10$ kpc, the "relativity parameter" $2GM/c^2R\approx 10^{-9}<<1$, which shows that in this range of the physical parameters general relativistic effects can be neglected, and the dynamics of the halos can be described with a very good approximation in the framework of Newtonian physics. Hence stable configurations are possible, and the halo would not collapse to a black hole, ending in a singular state. The collapse ends when the interaction energy equals the gravitational energy,
\be
\left|E_{grav}\right|=E_{int},
\ee
a condition which allows the approximate determination of the radius of the stable configuration as $R_{st}=\sqrt{C_{int}/C_{grav}}$, a result which is consistent with the radius of the static configuration given by Eq.~(\ref{Rst}) obtained by solving the full set of dynamical evolution equations. In the case of the gravitationally confined Bose-Einstein Condensates the interaction energy plays the same role as the electron and neutron quantum degenerate pressures in the cases of the white dwarfs and of the neutron stars, and it supports the dark matter halo against gravitational collapse.

 The total energy $E$ of the collapsed Bose-Einstein Condensates dark matter halo  can be written as $E = E_{zp} + E_{int} + E_{grav}$,
where $E_{zp}$, $E_{int}$ and $E_{grav}$  are the zero point kinetic energy, the interaction energy, and the gravitational
energy, respectively. In the following we will consider an approximate estimate of the energy, by following the approach considered in \cite{Pet}.  The kinetic energy per particle is $\hbar ^2/2m_{\chi}R_{st}^2$, and therefore the
total kinetic energy of the system is given by $E_{kin} = N\hbar ^2/2m_{\chi}R_{st}^2$,  where $N$ is the total particle number. The interaction energy
can be obtained as $E_{int} = (1/2) (N^2/V )mu_0$, where $V$ is the volume of the condensate, while the gravitational potential energy is
$E_{grav} =- GM^2/R_{st}$ (for simplicity we neglect in the following the factor $3/4$ in the expression of the gravitational potential energy). Therefore the total energy of the condensate is given by
\be
E=N\frac{\hbar ^2}{2m_{\chi}R_{st}^2}+\frac{3}{2}N^2\frac{\hbar ^2a}{m_{\chi}R_{st}^3}-\frac{GM^2}{R_{st}}.
\ee

By taking into account the explicit expressions of the radius and mass of the static condensate, given by Eqs.~(\ref{rad}) and (\ref{mass1}), respectively, we obtain for the total energy the expression
\be
E=\frac{2 \sqrt{a} \hbar ^3  \left[4 \pi  \left(3-2 \pi ^2\right) a^2 \hbar ^2
   \rho _c+G m_{\chi}^4\right]}{\sqrt{ G^3 m_{\chi}^{15}}}\rho _c.
   \ee

For $a=0$ we have $E=0$, but this case is not relevant for the study of the condensate dark matter halos. If the parameters of the BEC dark matter halo satisfy the constraint
   \be\label{condu}
   210.351\frac{a^2\hbar ^2}{Gm_{\chi}^4}\rho _c>1,
   \ee
then the total energy $E$ of the halo satisfies the stability condition $E<0$. The condition given by Eq.~(\ref{condu}) can be reformulated as
\be
   3.50\times 10^{45}\times \frac{\left(a/10^{-7}\;{\rm cm}\right)^2}{\left(m_{\chi}/10^{-32}\;{\rm g}\right)^4}\times \frac{\rho _c}{10^{-24}\;{\rm g/cm^3}}>1,
   \ee
   and it is obvious that it is satisfied by most of the realistic BEC dark matter halo models.

\subsection{The frequency of the small oscillations about the equilibrium state}

 Eqs.~(\ref{eqosc1}) and (\ref{eqosc2}) allow the study of the frequency of the small oscillations about the equilibrium state of the condensed dark matter halos. Expanding the potential $U(R)$ to second order in $R-R_{st}$ we find
\be
U(R)=U\left(R_{st}\right)+\frac{1}{2}K_{eff}\left(R-R_{st}\right)^2,
\ee
where $K_{eff}=U''\left(R_{st}\right)$. The equation of motion of the perturbed dark matter halo is
\be
m_{eff}\ddot{R}+K_{eff}\left(R-R_{st}\right)=0,
\ee
and therefore the frequency $\omega $ of the small oscillations is obtained as
\be
\omega ^2=\frac{K_{eff}}{m_{eff}}=\frac{R_{st}^2U''\left(R_{st}\right)}{R_{st}^2m_{eff}}=\frac{ 3E_{int}\left(R_{st}\right)-E_{grav}\left(R_{st}\right)}{R_{st}^2m_{eff}}.
 \ee
 With the use of Eq.~(\ref{Rst}) for the equilibrium radius, and of the expression of $m_{eff}$, the oscillation frequency can be expressed as
\be
\omega ^2=\frac{14 C_{grav}^{5/2}}{9 \sqrt{3} C_{int}^{3/2} M}>0,
 \ee
 or
 \be
\omega ^{2}=\sqrt{\frac{2}{7}}\frac{320}{567}\pi ^{3/2}G^{5/2}\frac{m_{\chi
}^{3}}{u_{0}^{3/2}}M=\sqrt{\frac{2}{7}}\frac{40}{567}\sqrt{\frac{G^{5}m_{\chi }^{9}}{%
a^{3}\hbar ^{6}}}M>0.
 \ee
 Equivalently, we obtain
 \bea
 &&\omega =4.837\times 10^{-17}\times \nonumber\\
 &&\left(\frac{m_{\chi}}{10^{-32}\;{\rm g}}\right)^{9/4}\left(\frac{a}{10^{-7}\;{\rm cm}}\right)^{-3/4}\left(\frac{M}{10^6M_{\odot}}\right)^{1/2}\;{\rm s}^{-1}.\nonumber\\
 \eea

\section{Discussions and final remarks}\label{sect6}

In this paper we have analyzed a simple model for the collapse of BEC dark matter halos,
based on the dynamical properties of the Gross-Pitaevskii equation. The present
model does not include damping nor a microscopic mechanism
for particle ejection. The rotational effects have also been ignored, as well as the possible presence of vortices in the condensate. We
also do not include any decay mechanism of the vortices (or turbulence)
during the collapse, since these effects can be neglected because of the
rapid increase of the density. Starting from the general description of
a time dependent Bose-Einstein gravitationally confined condensate, we focus our attention on a simple limiting case,
in order to obtain  some intuitive understanding of the physical properties
of the gravitational collapse of the dark matter halos, and the formation of stable astrophysical systems.

In order to obtain a simple mathematical description of the collapse process we have used a variational approach, in which a trial form of the condensate wave function was adopted, with all the physical parameters of the dark matter halos assumed to have a $r/R(t)$ dependence.  With the help of the trial wave function, the equation of motion and dynamic properties of the time dependent condensate can be obtained from an effective time-dependent Lagrangian, which describes the time evolution of the condensate radius $R$. If the condensate wave function depends on one or more
parameters, the resulting Lagrangian functional yields approximate Lagrangian equations
of motion for these parameters. With the help of the trial wave function one minimizes the action with respect to the free parameter (the Rayleigh-Ritz method). The choice of the trial wave function is not unique, and different choices may lead to different results.  The precision of the method depends on the number of free trial parameters, and on how physically realistic the trial function is.
The continuity and Poisson equations can be solved exactly, and the density (square of the wave function) and the gravitational potential of the dark matter halo can be explicitly obtained in an analytical form, thus allowing a complete description of the dynamical evolution of the condensate. The motion of the collapsing dark matter halos can be described as the motion of a single point particle with mass $m_{eff}$ in the force field generated by the effective potential $U(R)$, which incorporates the effects of the zero point kinetic energy $E_{zp}$, of the interaction energy of the condensate particles $E_{int}$, and of the gravitational energy $E_{grav}$. The variational procedure allows us to express these energies as a function of the radius $R(t)$ only. However, their explicit functional form also depends on the scattering length $a$, the mass of the dark matter particles, and the total mass of the dark matter halo. The trial wave function depends on a single parameter $R(t)$, the time dependent radius of the condensate. The time-dependent density profile, as well as the gravitational potential, are similar to the first order approximations of the static density and gravitational potential profiles, given by Eqs.~(\ref{app1}) and (\ref{app2}), respectively, with the static radius of the condensate $R_{BE}$ substituted by the time dependent radius $R(t)$, and with a time-dependent central density $\rho _{BE}^{(c)}(t)$. The variational procedure used can be extended and significantly improved by the choice of a trial wave function depending on several physical parameters.
The adopted approach allows a complete exact analytical treatment of the gravitational collapse of the condensed dark matter halos. Other choices of the trial wave function, or the increase of the number of free parameters would require the extensive use of numerical methods for the integration of the evolution equations. The expected error in this variational approach may be of the order of a few percent, when compared to the exact numerical solution.

The study of the equation of motion of the collapsing condensate shows that the collapse process ends with the formation of a stable configuration,  with radius $R_{st}$ and mass $M_{st}$. The resulting  configuration can be of stellar nature, or of galactic nature, depending on the physical processes and the initial mass of the dark matter halo. During the cosmological evolution such a collapse process could have played an important role  in the formation of the galactic structure, and of the dark matter halos. On the other hand local perturbations of the condensed dark matter could lead to the formation of smaller mass condensate stars. At the end of the collapse the density distribution of the formed stable static structure is given by
\be\label{89}
\rho _{BE}(r)=\frac{15M_{st}}{8\pi}\frac{1}{R_{st}^3}\left(1-\frac{r^2}{R_{st}^2}\right)=\rho _{BE}^{(c)}\left(1-\frac{r^2}{R_{st}^2}\right),
\ee
where the central density $\rho _{BE}^{(c)}$ of the condensate is given by
\bea\label{93}
&&\rho _{BE}^{(c)}=\frac{15}{8\pi}\left(\frac{2}{63}\right)^{3/2}\frac{G^{3/2}m_{\chi}^{9/2}}{\hbar ^3a^{3/2}}M_{st}=3.141\times 10^{-27}\times \nonumber\\ &&\left(\frac{m_{\chi}}{10^{-32}\;{\rm g}}\right)^{9/2}\left(\frac{a}{10^{-7}\; {\rm cm}}\right)^{-3/2}\left(\frac{M_{st}}{10^6M_{\odot}}\right)\;{\rm g/cm^3}.\nonumber\\
\eea
Eq.~(\ref{93}) gives the mass-central density relation for stable gravitationally confined condensed astrophysical objects. This central density is of the same order of magnitude as the central dark matter density of galactic dark matter halos. On the other hand, Eq.~(\ref{89}) is consistent with Eq.~(\ref{app1}), which gives an approximate representation of the density profile of the static density profile of the condensed Bose-Einstein dark matter halos.  In the first order approximation the density profile of the static condensate is $\rho _{BE}(r)\approx \rho _{BE}^{(c)}\left[1 -
  (\pi ^2/6)(r^2/R_{BE}^2)\right]$. With the start of the collapse, the radius of the halo becomes time-dependent, $R_{BE}\rightarrow R(t)$, while the central density of the initial static halo changes in time as $\rho _{BE}^{(c)}\rightarrow M/R^3(t)$. This shows that the choice of the trial function for the time dependent case is consistent with the static case.

An interesting question is the possibility of formation of dark dense Bose-Einstein condensed stars, having astrophysical properties (mass and radius) similar to those of the standard neutron stars. The radius and the mass of the dark star are determined by the mass of the dark matter particle, and by the scattering length. The radius of a neutron star is of the order of $R_{NS}\approx 10^{6}$ cm. A scaling of the mass and of the scattering length of the form $a=\alpha \times 10^{-7}$ cm, $m_{\chi }=\beta \times 10^{-32}$ g will give a radius of the same order of magnitude as the radius of a neutron star if the coefficients $\alpha $ and $\beta $ satisfy the condition $\alpha /\beta ^3=10^{-15}$. For $\alpha =10^{-12}$, giving $a=10^{-19}$ cm, we obtain $\beta =10$, which implies a dark matter particle mass of $m_{\chi}\approx 10^{-31}\;{\rm g}\approx 55$ eV. On the other hand, for realistic dark matter densities the corresponding mass of the star will exceed the general relativistic stability limit. Hence the realistic description of the dark stars requires the inclusion of general relativistic effects in the study of their structure \cite{Harko3}.

The details of the collapse of the Bose-Einstein condensate, as well as the numerical values of the physical parameters of the resulting stable configuration are strongly dependent on the numerical values of the two parameters describing the physical properties of the condensate, the dark matter particle mass $m_{\chi}$, and the scattering length $a$. The numerical values of these physical quantities are poorly known. We have discussed a number of observational constraints (galactic radii and Bullet Cluster data) that provide some limits for $m_{\chi}$ and $a$. Within the framework of the Bose-Einstein condensed dark matter model these astrophysical constraints point towards a dark matter particle with mass in the range of meV to a few eV, and a scattering length of the order of $10^{-19}$ cm. However, in the present paper most of the numerical results are normalized for a scattering length of $10^{-7}$ cm, and a mass of the dark matter particle of the order of $10^{-32}$ g. By a simple scaling all the numerical values corresponding to other choices of $m_{\chi}$ and $a$ can be obtained easily.
We have also considered the stability properties of the stable dark matter halos with respect to small oscillations, and the oscillations frequencies of the halos have also been obtained. These results show that the stable configuration, formed from the collapse of the condensed dark matter halos are stable with respect to small perturbations.

A large number of astrophysical observations, including the flat galactic cores, or the constant density surfaces points towards the possibility that dark matter may exist in the Universe in the form of a Bose-Einstein condensate, and that this possibility cannot be excluded {\it a priori}. The confirmation of this hypothesis by further observations on both galactic and cosmological scales would lead to a major change in our understanding of the basic principles of cosmology and astrophysics. In the present papers we have developed some theoretical tools that can help in the better understanding of the structure formation in the presence of condensed dark matter.

\end{document}